\documentclass[a4paper,11pt]{article}
\usepackage{pos}

\title{Determination of CP-violating Higgs couplings with transversely-polarized beams at the ILC}

\author*[a,b]{Cheng Li}
\author[b,c]{Gudrid Moortgat-Pick}

\affiliation[a]{School of Science, Sun Yat-Sen University, Gongchang Road 66, 518107 Shenzhen, China}
\affiliation[b]{Deutsches Elektronen-Synchrotron DESY,
  Notkestr. 85, 22607 Hamburg, Germany}

\affiliation[c]{II. Institut f{\"u}r Theoretische Physik\\  Universit{\"a}t Hamburg, Luruper Chaussee 149, 22761 Hamburg, Germany}

\emailAdd{cheng.li@desy.de}
\emailAdd{gudrid.moortgat-pick@desy.de}

\abstract{We study possible CP-violation effects of the Higgs to $Z$-boson coupling
at a future $e^+ e^-$ collider, e.g. the International Linear Collider (ILC). We find that the azimuthal angular distribution of the muon pair, produced by $e^+ e^- \rightarrow H Z \rightarrow H \mu^+ \mu^-$, can be sensitive to such a CP-violation effect when we apply initial transversely polarized beams. Based on this angular distribution, we construct a CP sensitive asymmetry and obtain this asymmetry by \texttt{Whizard} simulation. By comparing the SM prediction with 2$\sigma$ range of this asymmetry, we estimate the discovery limit of the CP-odd coupling in $HZZ$ interaction.}

\FullConference{The European Physical Society Conference on High Energy Physics (EPS-HEP2023)\\
 21-25 August 2023\\
Hamburg, Germany\\}

\begin{document}
\maketitle

\section{Introduction}
As we know from the Cosmic Microwave Background observation \cite{Planck:2018vyg}, the universe has much larger baryon number density than the SM prediction. In order to explain this discrepancy, the additional source of CP-violation has to be introduced according to Sakharov conditions for Baryogenesis \cite{Sakharov:1967dj}. In particular, one of the possible CP-violation source can be introduced in the Higgs sector, which is generated by the Two-Higgs-Doublet Model (2HDM) with spontaneous CP violation \cite{Lee:1973iz}. In this model, the CP-even 125 GeV Higgs boson gets the CP-odd admixture and the Higgs to fermions couplings are CP-violating. So far, such a Baryogenesis and CP-violation effect has been exploited by \cite{Bahl:2022yrs} via $Htt$ interaction at LHC , and via $H\tau\tau$ at Higgs factories \cite{Ge:2020mcl}. 

Furthermore, the tree-level CP-violating Higgs to fermions couplings can contribute a CP-violating Higgs to gauge bosons couplings at one-loop level. So far the LHC experiments, ATLAS \cite{ATLAS:2023mqy} and CMS \cite{CMS:2017len,CMS:2019jdw} perform the measurement for the $HVV$ coupling via $H\rightarrow 4\ell$ decay, and yield a limit of CP-odd $HVV$ coupling, which can be interpreted as $\left({\Tilde{c}_{AZZ}} \right)_\text{CMS} \sim [-1.65, 0.63]$ and $(\Tilde{c}_{AZZ})_\text{ATLAS}\sim [-1.2, 1.75]$. On the other hand, one can probe directly the CP-violating $HVV$ couplings at $e^+ e^-$ colliders (CEPC~\cite{Sha:2022bkt} and ILC), since a CP-violating $HVV$ coupling can lead to deviations in the Higgs strahlung process at $e^+ e^-$ colliders. In addition, the initial beams polarization can be applied at the ILC \cite{Moortgat-Pick:2005jsx}, where particularly the transverse polarization can be used to construct new CP sensitive observables \cite{Moortgat-Pick:2005jsx,Biswal:2009ar}. 

In this work, we will focus on the Higgs strahlung process at the ILC with a center of mass energy of 250 GeV and apply transversely polarized electron-positron beams. In such a case, we will study the azimuthal angular distribution of the muon pair, produced by the $Z$ decay, and construct the naive T-odd observable to probe the CP-odd $HZZ$ coupling. Eventually, we try to determine the discovery limit of the CP-odd $HZZ$ coupling by comparing with the SM CP-conserving 2$\sigma$ limit.
\section{Theoretical framework}
\subsection{CP-violation in $HZZ$ coupling and EFT interpretation}
Concerning the Higgs to $Z$ boson interaction, we have the following effective lagrangian \cite{Artoisenet:2013puc}:
 \begin{align}
                        \mathcal{L}_\text{EFF}&=c_\text{SM} \,Z_\mu Z^\mu H +  \frac{{c}_{HZZ}}{v}Z_{\mu\nu}{Z}^{\mu\nu}H +  \frac{\tilde{c}_{AZZ}}{v}Z_{\mu\nu}\tilde{Z}^{\mu\nu}H \\
         &=\kappa_\text{SM} \cos\xi_{CP} \,Z_\mu Z^\mu H +  \frac{{\kappa}_{HZZ}}{v}\cos\xi_{CP} Z_{\mu\nu}{Z}^{\mu\nu}H +  \frac{\tilde{\kappa}_{AZZ}}{v} \sin\xi_{CP} Z_{\mu\nu}\tilde{Z}^{\mu\nu}H,\label{eq:effl2}
                    \end{align}
where the CP-violation angle is parameterized by $\xi_{CP}$ and $\tilde{c}_{AZZ}$ denotes the loop-induced CP-odd coupling. In this case, the total cross-section with $HZZ$ interaction is the linear combinations of the following three terms
\begin{equation}
    \sigma_\text{tot} = |c_\mathrm{SM}|^2\sigma_\mathrm{SM} + |c_{HZZ}|^2\sigma_{HZZ}+|\tilde{c}_{AZZ}|^2 \sigma_{AZZ}.
\end{equation}
In this study, we ignore the CP-even operator $c_{HZZ}$ \cite{Li:2023hsr}. Furthermore, we can define a cross-section fraction \cite{Anderson:2013afp} to parameterize the CP-violation:
\begin{equation}
    f_{CPV} = \frac{|\tilde{c}_{AZZ}|^2 \sigma_{AZZ}}{|c_\mathrm{SM}|^2\sigma_\mathrm{SM} + |\tilde{c}_{AZZ}|^2 \sigma_{AZZ}},
\end{equation}
where the SM contribution is fixed by setting $|c_\mathrm{SM}|=1$.
\subsection{The scattering amplitude with polarized electron-positron beams}
\begin{figure}[h]
    \centering
    \includegraphics[width= .3\linewidth]{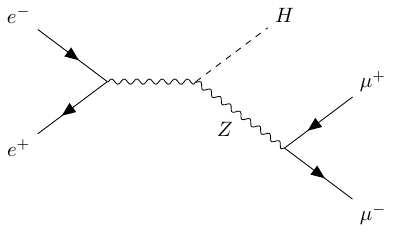}
    \caption{}
    \label{fig:diageezh}
\end{figure}

Concerning the polarization of the initial electron and positron beams, one can define a projection operator, which is given by:
\begin{equation}
\small
    \frac{1}{2}(1 - \boldsymbol{P}\cdot \boldsymbol{\sigma}) =\frac{1}{2}(\delta_{\lambda\lambda'}-P^a\sigma^a_{\lambda\lambda'}) = \frac{1}{2}\begin{pmatrix}
        1- P_3& P_1- iP_2\\
        P_1 + iP_2& 1+P_3
    \end{pmatrix} = \frac{1}{2}\begin{pmatrix}
                1-f\cos\theta_P& f\sin\theta_P e^{-i\phi_P}\\
                f\sin\theta_P e^{i\phi_P}&  1+f\cos\theta_P
            \end{pmatrix},
    \label{eq:polmat}
\end{equation}
where $f$ is the polarization fraction. The Higgs strahlung $e^+ e^- \rightarrow ZH$ is the dominant Higgs production process at $e^+ ~e^-$ collider. By taking the polarization of the initial beams into account, the spin density matrix of the Higgs strahlung $\rho_{\lambda_r \lambda_u}$ can be derived, see Fig.~\ref{fig:diageezh},
where the $\lambda_r, \lambda_u$ are the spin indices of the initial electron and positron. Particularly, we obtain the spin density matrix by applying the Bouchiat-Michel formula \cite{Bouchiat:1958yui, Li:2023hsr}:
\begin{align}
    u(p,\lambda_r')\Bar{u}(p,\lambda_r)=\frac{1}{2}\left(\delta_{\lambda_r\lambda_r'}+\gamma_5 s\!\!\!/^a \sigma^a_{\lambda_r\lambda_r'}\right)(p\!\!\!/+m),\\
    v(p,\lambda_u')\Bar{v}(p,\lambda_u)=\frac{1}{2}\left(\delta_{\lambda_u\lambda_u'}+\gamma_5 s\!\!\!/^a \sigma^a_{\lambda_u\lambda_u'}\right)(p\!\!\!/-m),
\end{align}
where $\sigma^a$ denotes the Pauli matrices, and the four-vector $s^a_\mu,~a=1,2,3$ are the three spin vectors, which are both orthogonal to each other and the corresponding four momentum.
Furthermore, we introduced the CP-odd operator for the $HZZ$ interaction as Eq.~\eqref{eq:effl2}, and obtain the following form of the total amplitude squared:
\begin{equation}
            \begin{split}
                    |\mathcal{M}|^2 &=(1-P_-^3P_+^3)( \cos^2\xi_{CP}\,\mathcal{A}_1^\text{CP-inv} + {\sin2\xi_{CP}\,\mathcal{A}^\text{CP-mix}} + \sin^2\xi_{CP}\,\mathcal{A}_2^\text{CP-inv})\\
                    &+(P_-^3-P_+^3)(\cos^2\xi_{CP}\,\mathcal{B}_1^\text{CP-inv} + {\sin2\xi_{CP}\,\mathcal{B}^\text{CP-mix}} + \sin^2\xi_{CP}\,\mathcal{B}_2^\text{CP-inv})\\
                    &+\sum_{mn}^{1,2}P_-^m P_+^n\left(\cos^2\xi_{CP}\,{\mathcal{C}^{mn}_1}^\text{CP-inv} + {\sin2\xi_{CP}\,\mathcal{C}_{mn}^\text{CP-mix}} + \sin^2\xi_{CP}\,{\mathcal{C}^{mn}_2}^\text{CP-inv} \right).
            \end{split}
            \label{eq:ampsqbsm}
            \end{equation}
Note that, the CP-invariant parts with $\cos^2\xi_{CP}$ and $\sin^2\xi_{CP}$ are both CP-conserving, while the CP-mixing terms with $\sin2\xi_{CP}$ violate the CP symmetry. According to the analytical calculation, 
via the transversely polarized terms the contributions of the CP-violating terms $C^{\text{CP-mix}}_{mn}$ can be extracted by another triple-product, which is given by
\begin{equation}
    \mathcal{C}^{mn}_\text{CP-mix} \propto \epsilon_{\mu\nu\rho\sigma}[({p}_{e^-} +{p}_{e^+})^\mu p_H^\nu p_{\mu^-}^\rho s_{e^-}^\sigma ]\propto(\vec{p}_H\times\vec{p}_{\mu^-})\cdot \vec{s}_{e^-}\propto(\vec{p}_{\mu^+}\times\vec{p}_{\mu^-})\cdot \vec{s}_{e^-}.
\end{equation}
The triple-product is the azimuthal-angle difference between the $\mu^+\mu^-$ plane and the spin of the electron. Therefore, defining the orientation of the azimuthal plane by fixing the direction of electron transverse polarization, the $C^{\text{CP-mix}}_{mn}$ depends directly on the azimuthal angle of the final state $\mu^-$.

\subsection{The asymmetry}
Regarding the azimuthal angular distribution of the muon plane, one can define an observable which sensitive to the CP-mixing effect: 
\begin{equation}
    \mathcal{O}_{CP} = \eta_{H} \sin 2 \phi_{\mu^-}.
\end{equation}
Hence, this asymmetry can be given by:
\begin{equation}
        \mathcal{A}_{CP} = \frac{1}{\sigma_\text{tot}} \int\operatorname{sgn}(\mathcal{O}_{CP}){d\sigma} = \frac{N(\mathcal{O}_{CP}<0)-N(\mathcal{O}_{CP}>0)}{N(\mathcal{O}_{CP}<0)+N(\mathcal{O}_{CP}>0)},
            \label{eq:asyN}
\end{equation}
where $N$ denotes the corresponding number of events. 
Since the SM is CP conserving for the neutral current, the SM background for this asymmetry is negligible. However, the number of events fluctuates statistically leading to the uncertainty of this asymmetry, which is based on a binomial distribution and given by:
\begin{equation}
    \Delta\mathcal{A}_{CP} = \sqrt{\frac{1-\mathcal{A}_{CP}^2}{N_\mathrm{tot}}}.
        \label{eq:delasy}
\end{equation}
\section{Analytical results}
By taking the uncertainties of the asymmetry into account, one can potentially distinguish the CP-mixing cases from the SM case with a given integrated luminosity and derive the unique CP-violation effect. For the SM CP-conserving case, the uncertainties of the asymmetries for both 500~fb$^{-1}$ and 2000~fb$^{-1}$ are given by
\begin{equation}
    \mathcal{A}_{CP}(500~\mathrm{fb}^{-1}) \approx (0\pm 1.58) \%,\qquad \mathcal{A}_{CP}(2000~\mathrm{fb}^{-1}) \approx (0\pm 0.79) \%.
    \label{eq:asysm}
\end{equation} 

\begin{figure}[h]
    \centering
    \includegraphics[width=.45\linewidth]{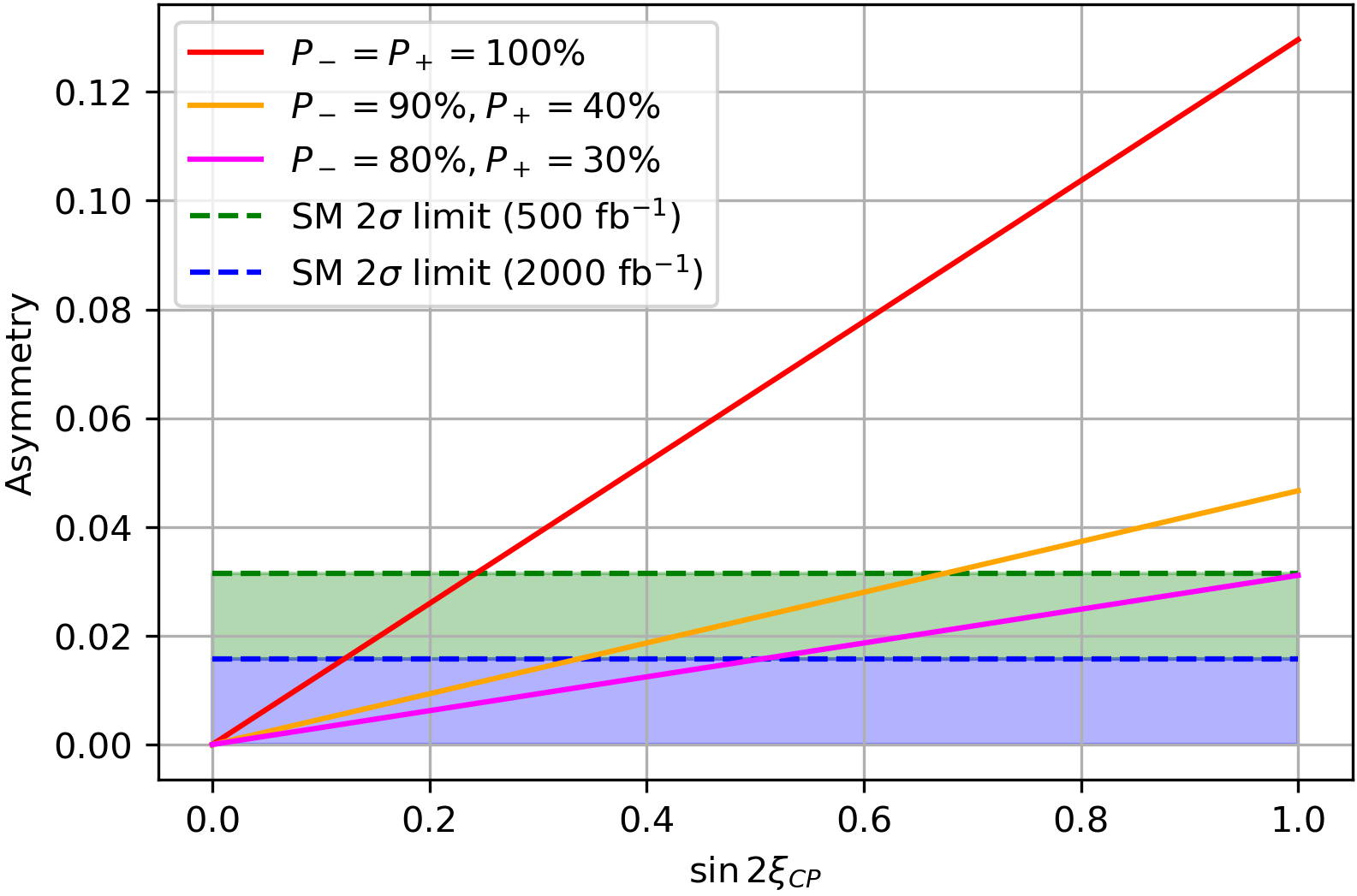}~~\includegraphics[width=.4\linewidth]{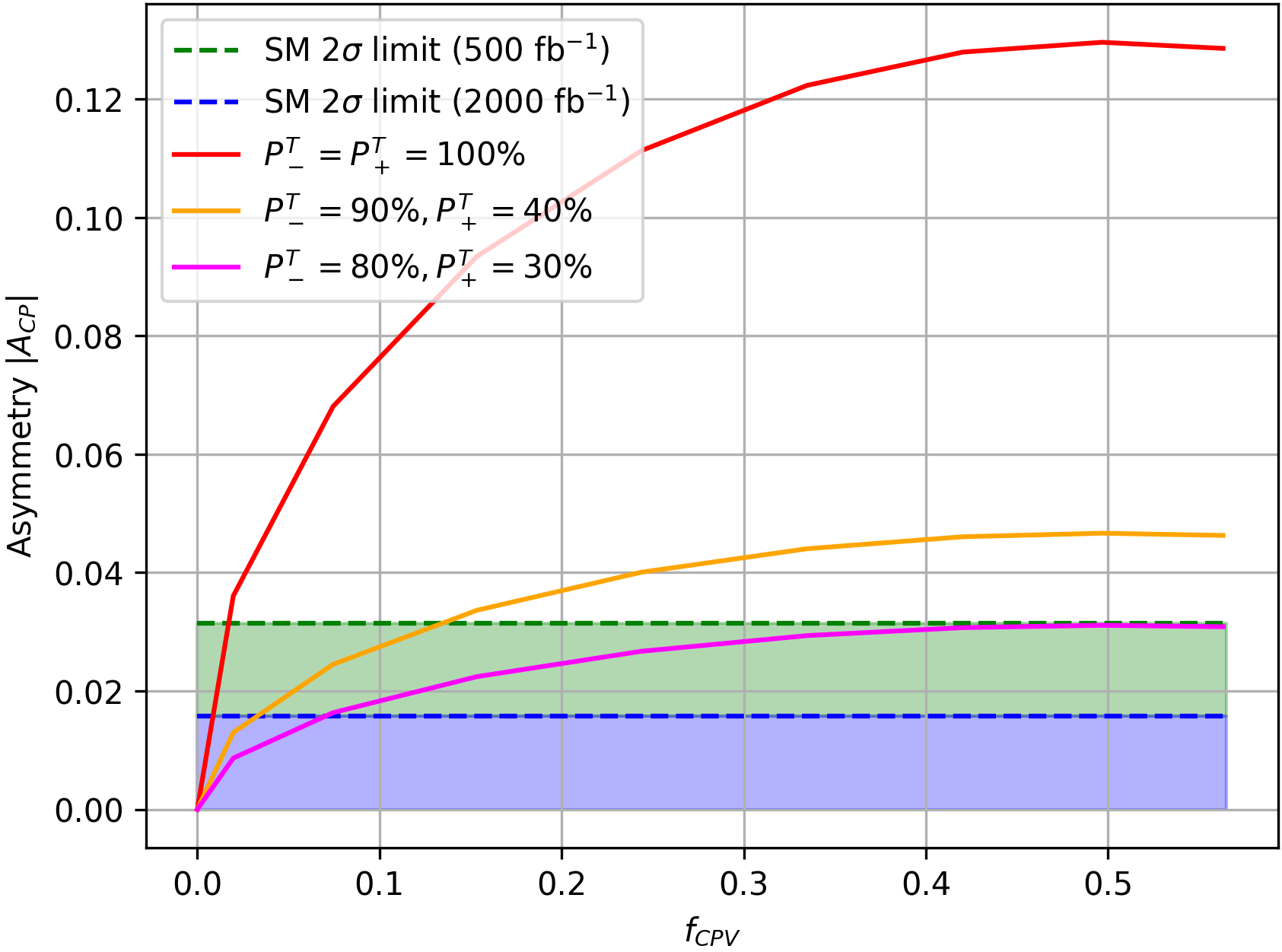}
    \caption{The analytical results of the asymmetries from Eq.~\eqref{eq:asyN} where the uncertainties of the asymmetries are taken from Eq.~\eqref{eq:delasy}. Left panel: varying $|\sin2\xi_{CP}|$ and fixed total cross section $\sigma_\mathrm{tot}=\sigma_\mathrm{SM}$; right panel: varying CP-odd coupling and fixing $c_\mathrm{SM}=1$. The red solid line corresponds to the completely polarized beams $(P^T_-, P^T_+) = (100\%, 100\%)$, while the orange line and magenta line demonstrate the asymmetries with $(P^T_-, P^T_+) = (90\%, 40\%)$ polarized beams and $(P^T_-, P^T_+) = (80\%, 30\%)$ polarized beams respectively. The blue and green region are the region below the 2$\sigma$ limit of SM CP-conserving case for 500~fb$^{-1}$ and $2000~\mathrm{fb}^{-1}$ respectively. }
    \label{fig:asymmetry}
\end{figure}

Thus, we vary the CP-mixing angles from the CP-conserving case $|\sin2\xi_{CP}| = 0$ to the maximal CP-mixing case $|\sin2\xi_{CP}|  = 1$, and present the asymmetries in Fig.~\ref{fig:asymmetry}, where we still fix the total cross section $\sigma_\mathrm{tot}=\sigma_\mathrm{SM}$ as used before. 

As we can see in the figure, the CP-conserving case with $|\sin2\xi_{CP}| = 0$ shows the vanishing asymmetry $\mathcal{A}_{CP}$, while this CP-sensitive asymmetry would be enhanced by increasing $|\sin2\xi_{CP}|$. By comparing with the SM results and its 2$\sigma$ region in Fig.~\ref{fig:asymmetry}, the transversely polarized beams configuration $(P^T_-, P^T_+) = (80\%, 30\%)$ cannot generate a large enough asymmetry $\mathcal{A}_{CP}$, where the $\mathcal{A}_{CP}(\sin2\xi_{CP}=1)$ is still within the 2$\sigma$ range of 500~fb$^{-1}$. However, the asymmetries for $|\sin2\xi_{CP}|\gtrsim0.5$ are above the blue region for 2000~fb$^{-1}$, which can be distinguished from the SM CP-conserving case at 95\% confidence level. Furthermore, we can use the polarization configuration of transversely polarized beams with $(P^T_-, P^T_+) = (90\%, 40\%)$, which is the best polarization fraction for the electron and positron beams expected by the current set-up \cite{Moortgat-Pick:2005jsx}. In this case, the limit of $|\sin2\xi_{CP}|$, where the asymmetry $\mathcal{A}_{CP}$ can be distinguished from the SM CP-conserving case, can be improved by the increment of the polarization fraction.


On the other hand, we can just fix the SM tree-level contribution by $c_\mathrm{SM}=1$, and vary the $\tilde{c}_{AZZ}$ individually, and  present the results of asymmetry $\mathcal{A}_{CP}$ in right panel of Fig.~\ref{fig:asymmetry}. This figure demonstrates that the asymmetry $\mathcal{A}_{CP}$ would reach a maximum for $f_{CPV}\approx 0.5$, where the CP-odd and CP-even interaction have equal contributions to the total cross-section. Note that, the maximum values of $\mathcal{A}_{CP}$ can be decreased by smaller transverse polarization fraction. The configuration with $(P^T_-, P^T_+) = (80\%, 30\%)$ polarized beams and the luminosity of 500~fb$^{-1}$ is still insufficient to determine the CP structure of the $HZZ$ interaction. However, the luminosity of 2000~fb$^{-1}$ improves this sensitivity significantly, and the fraction $f_{CPV}\sim0.1$ can be determined even by $(P^T_-, P^T_+) = (80\%, 30\%)$.

 \section{The discovery limits for CP-violation at ILC}
     

\begin{figure}
    \centering
    \includegraphics[width=.4\linewidth]{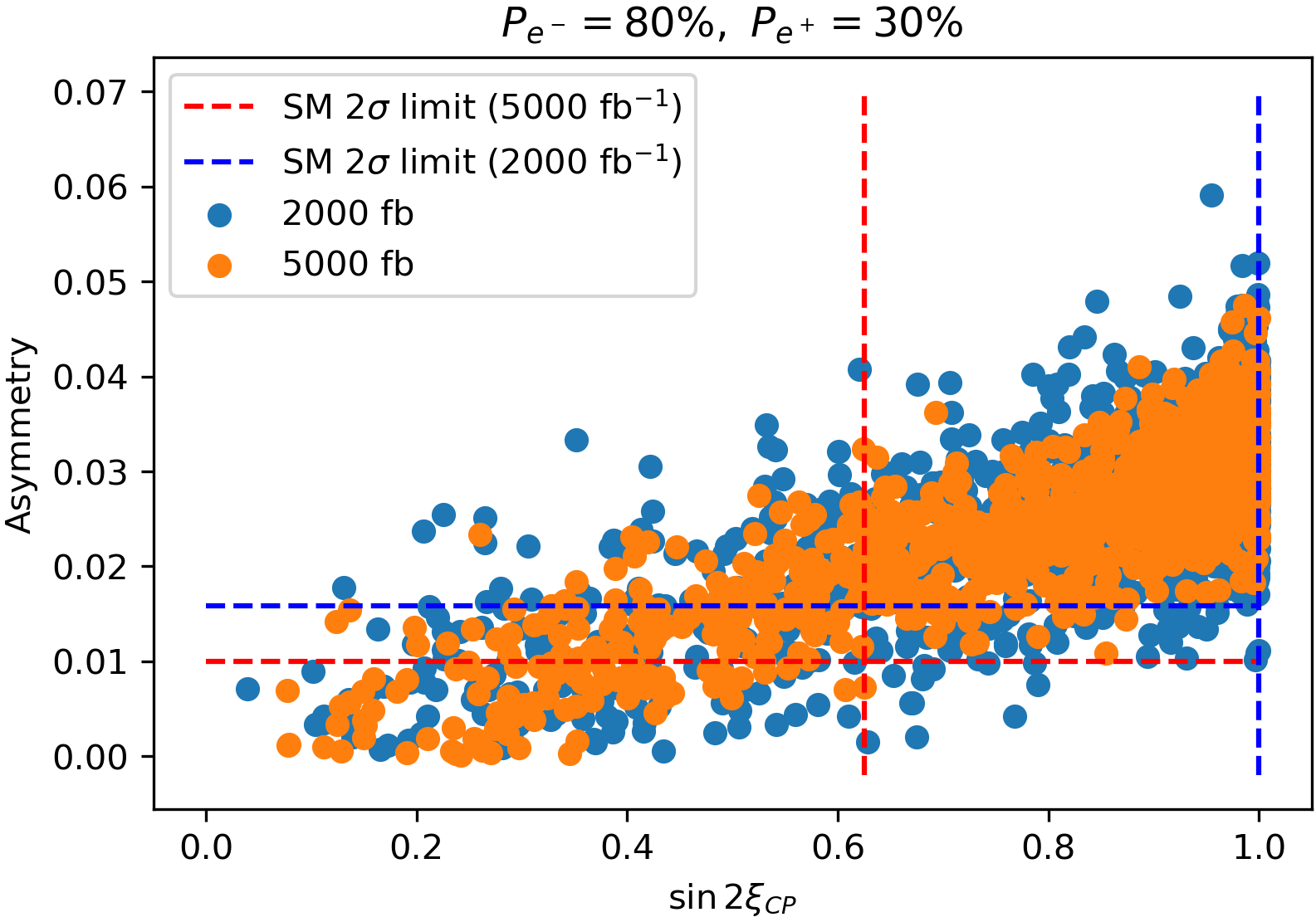}~~\includegraphics[width=.4\linewidth]{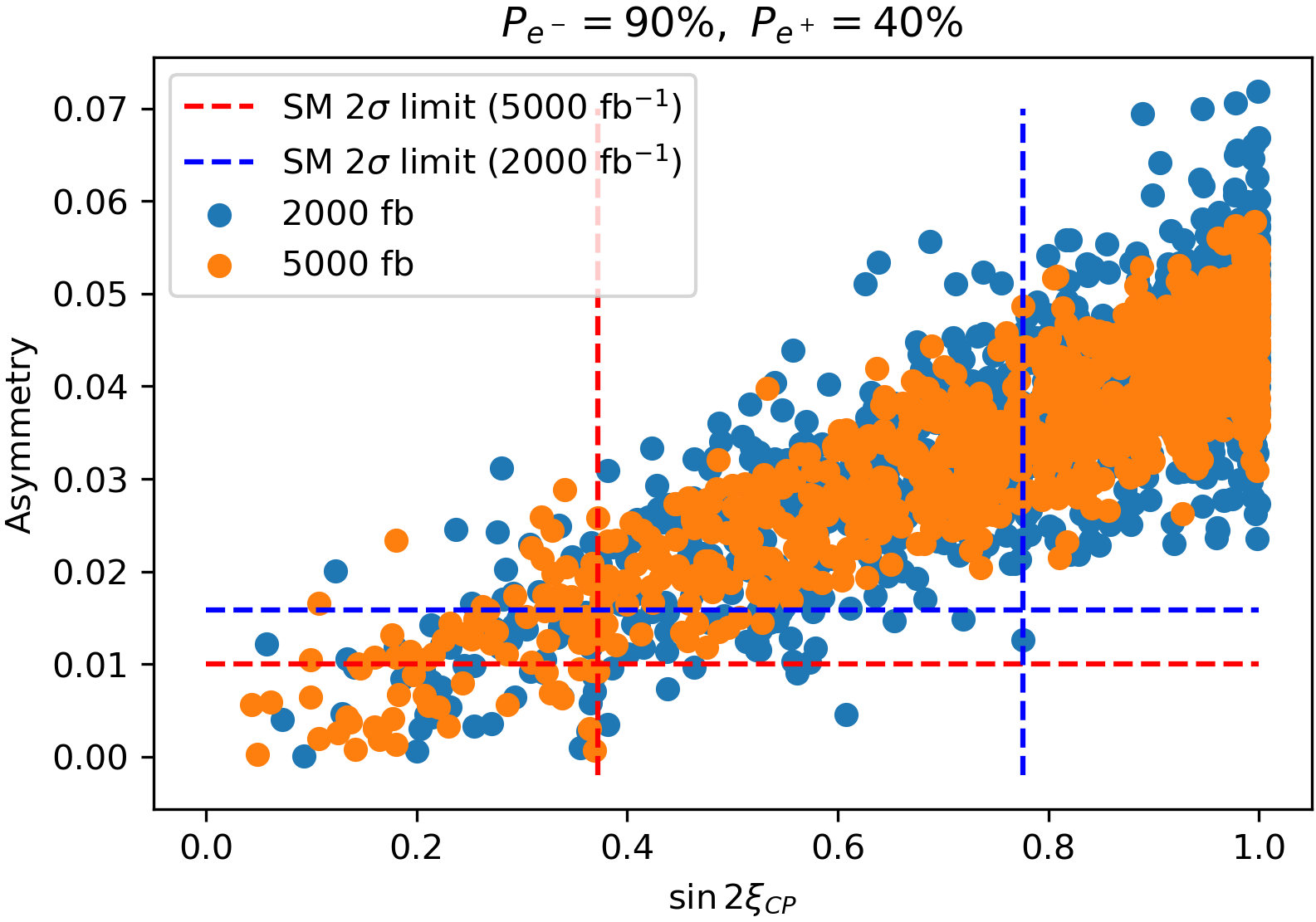}
    \caption{The Monte Carlo simulation results of the asymmetries generated by the $\texttt{Whizard}$, where the total cross-section is fixed to the SM tree-level value. The left panel corresponds to the polarization fraction $(P^T_-, P^T_+) = (80\%, 30\%)$, and the right panel corresponds to $(P^T_-, P^T_+) = (90\%, 40\%)$. For both figures, the results with luminosity 2000~fb$^{-1}$ are demonstrated by the blue points and 5000~fb$^{-1}$ for the orange points. In addition, the blue and orange horizontal dashed lines indicate the SM CP-conserving 2$\sigma$ limit.}
    \label{fig:cpmixingscan1}
\end{figure}
\begin{figure}[h]
    \centering
\includegraphics[width=.4\linewidth]{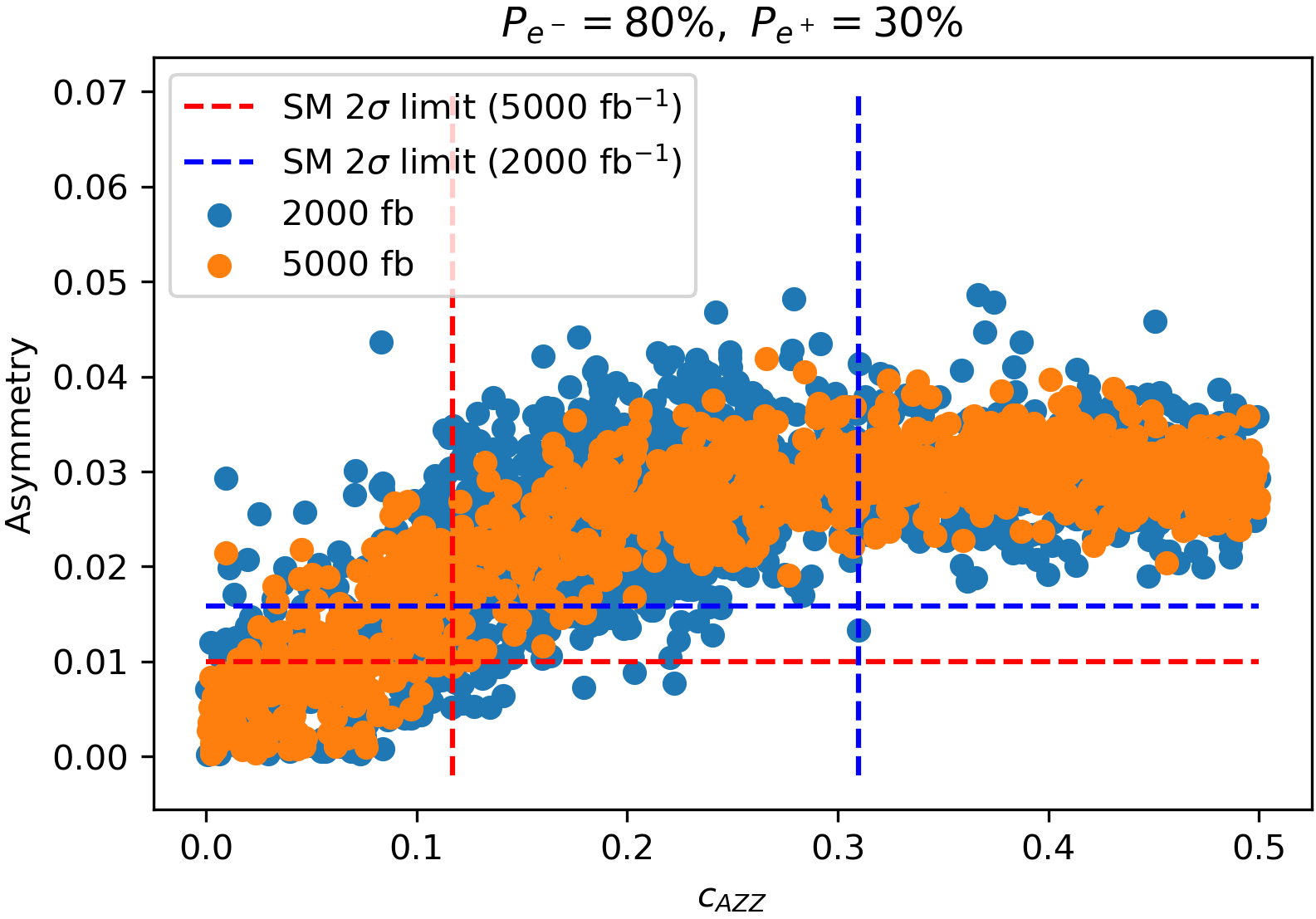}~~\includegraphics[width=.4\linewidth]{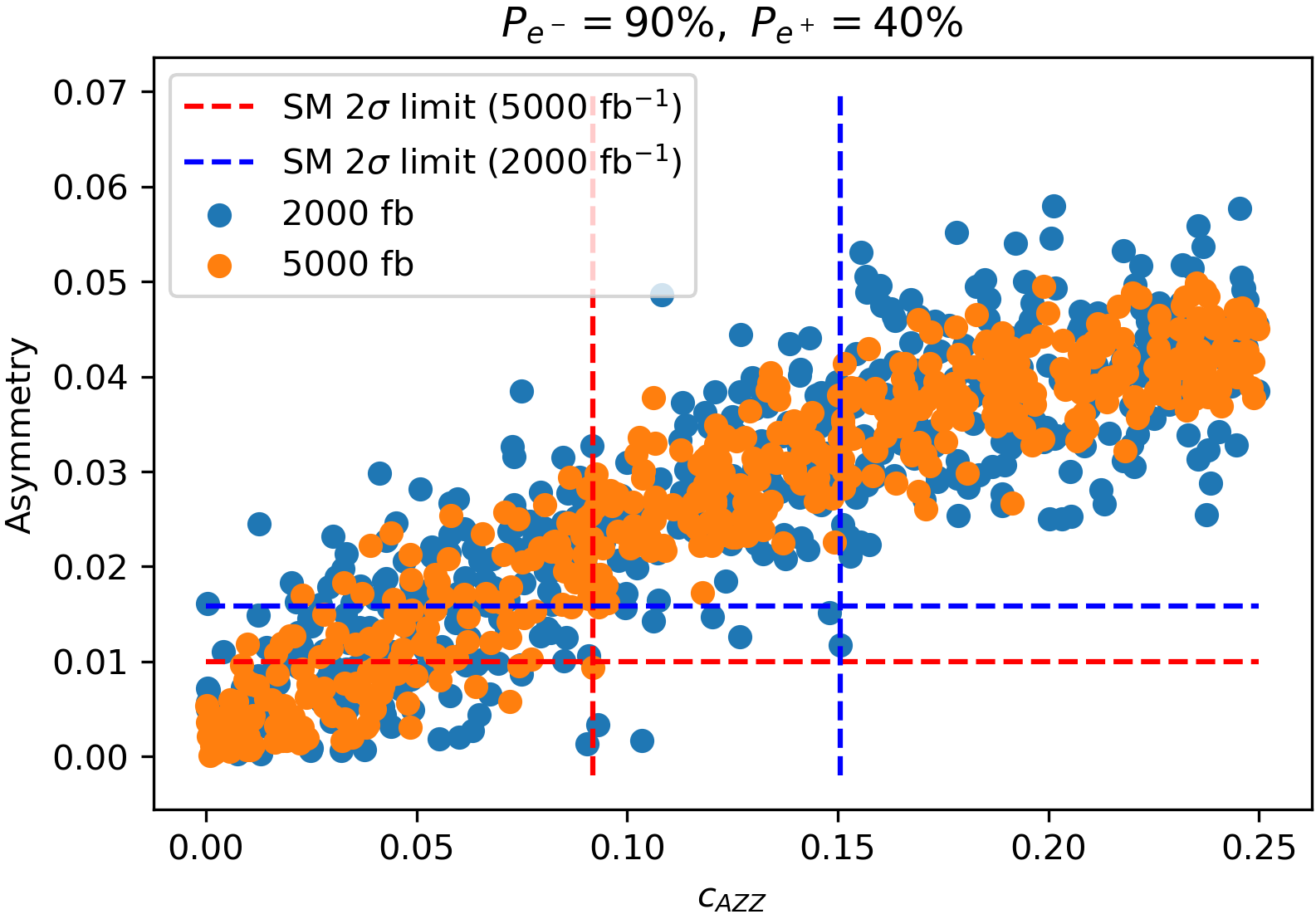}
    \caption{The Monte Carlo simulation results of the asymmetries generated by the $\texttt{Whizard}$, where the SM cross-section contribution is fixed. The left panel corresponds to the polarization fraction $(P^T_-, P^T_+) = (80\%, 30\%)$, and the right panel corresponds to $(P^T_-, P^T_+) = (90\%, 40\%)$. For both figures, the results with luminosity 2000~fb$^{-1}$ are demonstrated by the blue points and 5000~fb$^{-1}$ for the orange points. In addition, the blue and orange dashed horizontal lines indicate the SM CP-conserving 2$\sigma$ limit for the asymmetry.}
    \label{fig:fcpvcan}
\end{figure}
In order to estimate the discovery limit for CP-violating effects, one has to perform the Monte Carlo simulation and to determine the parameter space where the CP-odd observable is distinguishable from the SM CP-conserving prediction by 95\% C.L. In Fig~\ref{fig:cpmixingscan1}, we present the results of the asymmetry $\mathcal{A}_\mathcal{CP}$ with varying CP-mixing angle and the fluctuating total cross-section, which are generated by $\mathtt{Whizard}$. As we see in Fig.~\ref{fig:cpmixingscan1}, the asymmetry is consistent with the analytical calculation in Fig.~\ref{fig:asymmetry}, and the asymmetry shows the statistical fluctuation in Fig.~\ref{fig:cpmixingscan1} for certain $|\sin2\xi_{CP}|$ and for the integral luminosity $2000~\mathrm{fb}^{-1}$ and $5000~\mathrm{fb}^{-1}$. Since the $500~\mathrm{fb}^{-1}$ is insufficient to determine the CP-violation effect based on the previous discussion, we do not present the results with $500~\mathrm{fb}^{-1}$ for the Monte Carlo simulation. By comparing the SM the $2\sigma$ limit lines, we determine that 2$\sigma$ discovered region with $5000~\mathrm{fb}^{-1}$ is around $|\sin2\xi_{CP}|\gtrsim0.6$, which is presented as the vertical line in the figure, and CP-violation effects can be distinguished from the SM prediction in this region. However, the blue points with maximal CP-violation effect $|\sin2\xi_{CP}|=1$ can be still below SM $2\sigma$ limit. This means one cannot distinguish the CP-violation effect with the integrate luminosity of $2000~\mathrm{fb}^{-1}$ and $(P^T_{e^-} = 80\%,P^T_{e^+} = 30\%)$ initial polarization, only taking the statistical uncertainties into account.

In the right panel of Fig~.\ref{fig:cpmixingscan1}, we present the results applying $(P^T_{e^-} = 90\%,P^T_{e^+} = 40\%)$ initial polarization. In this case, the discovered region of the CP-violation effect is enlarged by higher polarization degrees, where the discovery limit for $5000~\mathrm{fb}^{-1}$ is around $|\sin2\xi_{CP}|\gtrsim0.4$. In addition, the integrated luminosity of $2000~\mathrm{fb}^{-1}$ is able to determine the CP-violation effect with $|\sin2\xi_{CP}|\gtrsim0.8$.

Furthermore, we present the results with fixed the SM tree-level contribution but increasing the CP-odd coupling in Fig~\ref{fig:fcpvcan}. 
We can observe that the variation of the asymmetry $\mathcal{A}_{CP}$ is basically the same as the analytical results in Fig.~\ref{fig:asymmetry} but with additional statistical fluctuation. As the same as the discussion for previous scenario, we could read out the discovery limit of CP-odd coupling $\tilde{c}_{AZZ}$ from the vertical dashed lines in both panels. 

In order to summarize all the determination results, we present all discovery limits for the CP-violation effect for the scenarios we discussed in Tab.~\ref{tab:sum}.
\begin{table}[h]
    \centering
                 \begin{tabular}{cccc}
                       \hline
                          $(P^T_{e^-},P^T_{e^+})$& Luminosity [fb$^{-1}$]& $\sin 2\xi_{CP}$ limit& $c_{AZZ}$ limit  \\
                          \hline
                          (80\%, 30\%)&    2000 & -& [-0.31,0.31]\\
                        (80\%, 30\%)&    5000 & [-0.62,0.62]& [-0.12,0.12]\\
                                  (90\%, 40\%)&    2000 & [-0.79,0.79]& [-0.15,0.15]\\
                        (90\%, 40\%)&    5000 & [-0.39,0.39]& [-0.09,0.09]\\
                        \hline
                
                    \end{tabular} 
    \caption{The summary table for the polarization fraction $(P^T_{e^-},P^T_{e^+}) = (80\%, 30\%)$ and $(90 \%, 40\%)$, and the integrated luminosity of 2000~fb$^{-1}$ and 5000~fb$^{-1}$. The limit of $\sin2\xi_{CP}$ can be determined by assuming the total cross-section is fixed, and the limit of individual $\tilde{c}_{AZZ}$ is determined by fixing to the SM tree-level cross-section.}
    \label{tab:sum}
\end{table}
\section{Conclusion}
In this work, we investigate the CP-violation effects in the process $e^+ e^- \rightarrow HZ\rightarrow H \mu^+ \mu^-$ with initial transversely polarized beams. Based on the azimuthal angular distribution of the final state muons, we constructed a CP sensitive observable and exploited its dependence on the CP-violation parameters. Furthermore, we could determine discovery limit of the CP-violation effect by comparing with the 2$\sigma$ limit of the SM CP-conserving prediction. Finally, we obtained the limit of the CP-odd coupling $|\tilde{c}_{AZZ}|\lesssim 0.09$ with 5000~fb$^{-1}$ and $(P^T_{e^-},P^T_{e^+}) = (90 \%, 40\%)$ polarization, which is potentially better than the unpolarized result $|\tilde{c}_{AZZ}|\lesssim 0.15$ with 5600~fb$^{-1}$ in \cite{Sha:2022bkt}. Overall, these CP-odd coupling limit can be strongly improved at $e^+e^-$ collider with initial beams polarization compared to current LHC measurement.


\bibliographystyle{JHEP}   
\bibliography{ref}

\end{document}